\documentclass[a4paper,nohyper]{JHEP3}

\usepackage{epsfig,multicol}

\title{Lovelock type gravity and small black holes in heterotic
  string theory}

\author{Predrag Prester\\
Max Planck Institute for Gravitational Physics (Albert Einstein Institute)\\
Am M\"{u}hlenberg 1, D-14476 Golm, Germany\\ \\
Theoretical Physics Department, Faculty of Natural Sciences and
Mathematics\\ p.p. 331, HR-10002 Zagreb, Croatia\\ \\
E-mail: \email{pprester@phy.hr}}

\preprint{AEI-2005-181}

\abstract{We analyze near horizon behavior of small D-dimensional 
2-charge black holes by modifying tree level effective action of 
heterotic string with all extended Gauss-Bonnet densities. We show 
that there is a nontrivial and unique choice of parameters, 
independent of $D$, for which the black hole entropy in any dimension 
is given by $4\pi\sqrt{nw}$, which is exactly the statistical entropy
of 1/2-BPS states of heterotic string compactified on 
$T^{9-D}\times S^1$ with momentum $n$ and winding $w$. This extends 
the results of Sen [{\em JHEP} {\bf 0507} (2005) 073] to all 
dimensions. We also show that our Lovelock type action belongs to the 
more general class of actions sharing the simmilar behaviour on the 
$AdS_2\times S^{D-2}$ near horizon geometry.}

\begin{document}

\section{Introduction}

Recently black holes in heterotic string theory had attracted a lot of
attention\footnote{A overview of recent results for black holes in 
string theory is given in \cite{witt0511}.}. Special class are 
2-charge small black holes. On the string side these black holes
should correspond to perturbative half-BPS states of heterotic string 
compactified on $T^{9-D}\times S^1$, with momentum and winding on 
$S^1$ equal to $n$ and $w$, respectively, for which one can easily 
calculate asymptotic expression ($n,w\gg1$) for the number of states 
\cite{DabHar89,DaGiHaRR90}. Logarithm (which is the entropy in 
microcanonical ensamble) is in the leading order given by
\begin{equation} \label{entropy}
S=4\pi\sqrt{nw}
\end{equation}
This result, obtained for a free string, due to supersymmetry remains
to be valid after switching on the string coupling $g_s$. Now, as the
string coupling is increased, at one point de Broglie-Compton 
wavelength $1/M$ becomes smaller then the corresponding Schwarzschild
radius $\ell_P^2 M \sim g_s^2 \alpha' M$, which should lead to 
formation of (extremal) black hole. This is a one way to argue that
elementary string states with mass large enough should describe black
holes \cite{thoft90,sussk93,SusUgl94,HorPol96}.

Indeed, exact black hole solutions of the low energy effective action
of heterotic string theory in the leading order in $\alpha'$ were 
found which decribe D-dimensional extremal black holes with
``correct'' quantum numbers (e.g., they have two electric charges 
proportional to $n$ and $w$) \cite{sen9411,peet95}. They are in some 
sense pathological having null singularities and zero horizon 
area\footnote{This is the reason why they are called small or 
microscopic.}. This implies vanishing Bekenstein-Hawking entropy which
is obviously in disagreement with the string result (\ref{entropy}).

To understand what is happening, one should go back to the derivation
of (\ref{entropy}) -- and to see that although it is perturbative in
string coupling, it is {\em nonperturbative} in $\alpha'$. This means
that on the gravity side one should start from the complete tree-level
(in string coupling) effective action which contains all $\alpha'$
higher-derivative corrections. This is also visible from the structure
of the solution in the leading order -- singularity of the horizon 
implies that one cannot neglect higher curvature terms (or treat them
as perturbation) in the efective
action near the horizon, as it is usually done for large black holes.
In fact, a priory all terms should be of the same importance. The
remarkable property of small black holes is that they give us some
information on the {\em complete} tree-level (in string coupling) 
effective action.

In \cite{dabh0409,DakaMa04,sen0411,HuMaRa04,sen0502,sen0504} it was
shown in $D=4$ that adding to the action just one type of the 
higher-derivative terms, obtained by supersymmetrizing square of the
Weyl tensor \cite{CaWiMo98,CaWiKaMo99,CaWiKaMo00}, one obtains that 
corrected black holes have regular horizon of $AdS_2\times S^2$ type,
for which generalised Wald entropy formula\footnote{Note that 
although Wald derivation demands existence of the bifurkate Killing
horizon, and so does not apply to extremal black holes, one can 
formally take the limit of extremality in the final formula.} 
\cite{wald93,IyWa94,JaKaMy94} gives a desired result (\ref{entropy}).
This result is at the same time exciting and  mysteriuos, because 
there is no apparent reason why should only terms quadratic in
curvature contribute to the entropy, with all higher-order terms 
somehow cancelling.\footnote{In \cite{KraLar0506,KraLar0508} an
explanation was presented based on anomalies induced by particular
Chern-Simons terms. However, it is not clear to us why only those
terms should contribute.} It is important to note that for the entropy 
one only needs behaviour of the solution near the horizon, so this
cancelation could just appear there (as a consequence of the
$AdS_2\times S^2$ geometry). Indeed, numerical extrapolations to 
the far-away region show that solution does not aproach to 
Schwarzschild solution but has oscillating behaviour connected with 
spurious degrees of freedom typicaly present in higher order gravity
theories \cite{sen0411,HuMaRa04}. This could suggest that other 
higher order terms become important away from horizon.

A natural question is what is happening in $D>4$? Unfortunately, it 
is imposible to perform the same analysis, as it is not known how to
supersymmetrize $R^2$-terms in the action. In lack of this, Sen
\cite{sen0505} took as a ``toy-model'' just the gravitational part, 
which is proportional to Gauss-Bonnet density\footnote{There is also a
term proportional to the Pontryagin density, but it vanishes 
identicaly in $AdS_2\times S^n$ background.}, and analysed 
near-horizon behaviour of the solution (for which he assumed 
$AdS_2\times S^{D-2}$ geometry). Although this action is not 
supersymmetric, surprisingly, Wald entropy formula again gave 
(\ref{entropy}), now in $D=4$ and $D=5$ (but not for $D\ge6$). Even
more surprisingly, in the recent paper \cite{sen0508}, 
it was shown that for the same type of the action, applied to the 
large class of 8-charge black holes in $D=4$, entropy, near
horizon metric, gauge field strengths and the axion-dilaton field are
identical to those obtained in \cite{CaWiMo9906,CaWiKaMo04} from a 
supersymmetric version of the theory based on squared Weyl tensor. 

In this paper we extend Sen's analysis of two-charge black holes to
any number of dimensions $D\ge4$. For the effective action near
the horizon we take obvious generalisation, i.e., we use extended
Gauss-Bonnet densities as higher-order terms in curvature
\cite{love71,love72}. These ``Lovelock type'' actions have several 
appealing properties, e.g, they are of the first order (no ghosts or 
spurious states \cite{zwibach85,zumino86}), have good boundary value
problem, and contain only finite number of terms. We perform near 
horizon analysis assuming $AdS_2\times S^{D-2}$ geometry and, using 
Wald formula, calculate entropy, which has a complicated dependance on 
$D$ and\footnote{$[x]$ denote integer part of $x$.} $[D/2]$ coupling 
constants\footnote{$[D/2]$ is the number of extended Gauss-Bonnet
terms in $D$ dimensions, including the Einstein term.} $\lambda_m$.
We show that there is a unique choice for $\lambda_m$ (independent
of $D$) which gives exactly the expression (\ref{entropy}) in 
{\em any} $D$. It should be emphasized that this is a nontrivial 
result, in the sence that to fix the entropy for $D$ black holes 
one has only $[D/2]$ free parameters to play with (or, in other words,
for each couple of dimensions enters only one parameter). This
result trivially extends to black holes with more electric 
charges, connected with heterotic string compactifications on
$M_D\times T^{10-D-k}\times (S^1)^k$.

\section{Effective action with extended Gauss-Bonnet terms}

We are interested in heterotic string compactified on 
$T^{9-D}\times S^1$, for which effective low energy action in the
leading order in string coupling can be written in the form
\begin{equation} \label{treeea}
S = \frac{1}{16\pi G_N} \int d^Dx \sqrt{-g} \,S \sum_{m=1} 
\alpha'^{m-1} \mathcal{L}_m
\end{equation}
where $S$ is the dilaton field, which is connected to the effective 
closed string coupling constant $g$ by $S=1/g^2$.

Leading order term in $\alpha'$ is given by \cite{sen0505}
\begin{equation} \label{alpha0}
\mathcal{L}_1 = R + S^{-2}(\nabla S)^2 - T^{-2}(\nabla T)^2 
- T^2 \left(F_{\mu\nu}^{(1)}\right)^2 
- T^{-2} \left(F_{\mu\nu}^{(2)}\right)^2
\end{equation}
where we assumed that all other fields are vanishing. In this order
exact half-BPS electricaly charged extremal black hole solutions in 
any $D$ were found \cite{peet95} which have the same quantum numbers 
as perturbative half-BPS string states (where two electric charges are
proportional to momentum and winding of the string along $S^1$). These
solutions have singular 
horizon (null singularity) with a vanishing area, on which effective
string coupling also vanishes. This properties are in contrast
with what one expects from string theory, which for example gives the
nonvanishing result for the entropy (\ref{entropy}).

It is obvious what is wrong in the above analysis. As the horizon is
singular, the curvature invariants (and some other fields like $S$) 
are also, which means that in the effective action (\ref{treeea}) one
cannot neglect higher-order terms which typicaly contain higher 
powers and/or derivatives of the Riemann tensor. In $D=4$ dimensions 
it was shown in \cite{dabh0409,DakaMa04,sen0411,HuMaRa04} that if one
adds a particular class of higher-derivative terms (obtained by 
supersymmetrization of the square of the Weyl tensor), corrections 
completely change the nature of singularity - one gets timelike 
singularity hidden behind a horizon with the finite area. Also, the
dilaton field $S$ becomes finite on the horizon, which means that 
effective string coupling is nonvanishing. Using Wald formula it was
shown that the entropy is equal to the string result (\ref{entropy}).
Now, the mystery is why other terms, which are known to be present in
the effective action (especially ones containing higher powers of
the Riemann tensor), are appearing to be irrelevant for the
entropy calculation.

One way to understand what is happening would be to make the same
analysis in higher dimensions. Unfortunately, for $D>4$ supersymmeric
version of the action containing curvature squared terms is not known.
In lack of this, in \cite{sen0505} Sen took as a toy model an action
obtained by adding just the Gauss-Bonnet term. Although this action is
not supersymmetric, from the near horizon analysis he obtained that
the entropy is again given by (\ref{entropy}), but only in $D=4,5$.
Now, the interesting thing is that in $D=6$ a next extended 
Gauss-Bonnet term is present, so the natural question to ask is what
is happening if we include in the action all extended Gauss-Bonnet
terms. That is the main subject of this paper.

We propose to analyse the actions of the Lovelock type 
where higher order terms in $\alpha'$ in (\ref{treeea}) are given by 
the extended Gauss-Bonnet densities \cite{love71,love72}
\begin{equation}\label{lgbm}
\mathcal{L}_m = \lambda_m \mathcal{L}^{GB}_m = \frac{\lambda_m}{2^{m}}
\, \delta_{\mu_1\nu_1\ldots\mu_m\nu_m}^{\rho_1\sigma_1\ldots
\rho_m\sigma_m} \, {R^{\mu_1\nu_1}}_{\rho_1\sigma_1}\cdots
{R^{\mu_m\nu_m}}_{\rho_m\sigma_m}\;, \qquad m=2,\ldots,[D/2]
\end{equation}
where $\lambda_m$ are some (at the moment free) dimensionless 
parameters, $\delta_{\alpha_1\ldots\alpha_k}^{\beta_1\ldots\beta_k}$ 
is totally antisymmetric product of $k$ Kronecker deltas, normalized 
to take values 0 and $\pm 1$, $[x]$ denote integer part of $x$, and 
all greek indeces are running from 0 to $D-1$. Extended Gauss-Bonnet 
densities $\mathcal{L}^{GB}_m$ are in many respects generalisation of 
the Einstein term (note that $\mathcal{L}^{GB}_1=R$). Especially,
$m$-th term is topological in $D=2m$ dimensions. Also note that 
they identicaly vanish for $m>[D/2]$, so for any $D$ there is a finite 
number of terms in the action.

\section{Near horizon analysis}

We want to study solutions of the action given by
(\ref{treeea}--\ref{lgbm}) which should be deformations of
the exact small black hole solutions obtained in lowest order in
$\alpha'$. We do not know how to exactly solve equations of
motion, but we are primarly interested in the entropy which is given 
by the Wald formula \cite{wald93,IyWa94,JaKaMy94}
\begin{equation}\label{wald}
S = 2\pi \int_\mathcal{H} \hat{\epsilon} \, 
\frac{\partial\mathcal{L}}{\partial R_{\mu\nu\rho\sigma}}
\eta_{\mu\nu}\eta_{\rho\sigma}
\end{equation}
Important here is to notice that integration is done on the cross
section of the horizon $H$, so to calculate the entropy one only needs
to know a solution near the horizon.

Now, in \cite{sen0506} it was shown that symmetries of the horizon can
enormously simplify calculation of the entropy. In $D=4$ case it was
shown that near horizon geometry is of $AdS_2 \times S^2$ type, where
effect of $\alpha'$ corrections was to make radius of horizon
nonvanishing. Following \cite{sen0505} we conjecture that the same 
happens in $D>4$ so the near horizon geometry should be 
$AdS_2 \times S^{D-2}$. This implies that near the horizon fields have
the following form
\begin{eqnarray}
&& ds^2 = g_{\mu\nu}\, dx^\mu dx^\nu = v_1 \left( -x^2 dt^2 +
\frac{dx^2}{x^2} \right) + v_2\,d\Omega_{D-2}^2 \nonumber \\
&& S = u_S \nonumber \\
&& T = u_T \nonumber \\
&& F_{rt}^{(i)} = e_i \;, \qquad i=1,2 \label{horfie}
\end{eqnarray}
where $v_i$, $u_S$, $u_T$, $e_i$ are constants, and moreover that the
covariant derivatives of the scalar fields $S$ and $T$, the gauge
fields $F_{\mu\nu}^{(i)}$ and the Riemann tensor
$R_{\mu\nu\rho\sigma}$ vanish on the horizon $x=0$. This makes 
solving the equations of motions (EOM's) near the horizon (i.e., 
finding $v_i$, $u_S$, $u_T$ and $e_i$) very easy. One first defines
\begin{equation}\label{fuve}
f(\vec{u},\vec{v},\vec{e}) = \int_{S^{D-2}} \sqrt{-g}
\, \mathcal{L}
\end{equation}
where the integration is over $S^{D-2}$, and one uses (\ref{horfie}).
Equations of motion are near the horizon given by
\begin{equation}\label{eom}
\frac{\partial f}{\partial u_S}=0 \;,\qquad 
\frac{\partial f}{\partial u_T}=0 \;,\qquad
\frac{\partial f}{\partial v_1}=0 \;, \qquad
\frac{\partial f}{\partial v_2}=0
\end{equation}
Notice that configuration (\ref{horfie}) solves EOM's for gauge fields
identicaly on the horizon for any $e_i$. We also need to know electric
charges $q_i$. In \cite{sen0506} it was shown that they are given by
\begin{equation}\label{charge}
q_i = \frac{\partial f}{\partial e_i} \;,\qquad i=1,2
\end{equation}
We would also like to connect conserved charges (\ref{charge}) with
corresponding quantum numbers of half-BPS states of heterotic string,
which are momentum $n$ and winding $w$ around $S^1$. This is given by 
\cite{sen0508}
\begin{equation}\label{q-nw}
q_1 = \frac{2\,n}{\sqrt{\alpha'}} \;,\qquad 
q_2 = \frac{2\,w}{\sqrt{\alpha'}}
\end{equation}
It was shown in \cite{sen0506} that the entropy for the configuration
(\ref{horfie}) is given by
\begin{equation}
S = 2\pi \left( \sum_{i=1}^2 e_i \, q_i - f \right)
\end{equation}
For the actions of the type (\ref{treeea}) EOM for dilaton $S$ implies
that $f$ vanishes on-shell near the horizon, so we have just
\begin{equation}\label{ent}
S = 2\pi \sum_{i=1}^2 e_i \, q_i
\end{equation}

\section{Entropy of small black holes}

We now apply procedure from the previous section to analyse extremal
small black hole solutions in $D$ dimensions, with the $AdS_2\times
S^{D-2}$ horizon geometry, when the action is given by
(\ref{treeea}--\ref{lgbm}). First we need to calculate 
function $f$ (\ref{fuve}) using (\ref{horfie}). It was shown 
\cite{PP02} that for the metrics of the type
\begin{equation}\label{gsph}
ds^2 = \gamma_{ab}(x) dx^a dx^b + r(x)^2 d\Omega_{D-2}\;,\qquad
a,b=1,2
\end{equation}
the Gauss-Bonnet densities, integrated over the unit sphere $S^{D-2}$, 
give
\begin{eqnarray} \label{lmint}
\int_{S^{D-2}} \sqrt{-g}\,\mathcal{L}_m &=& - \Omega_{D-2} \lambda_m
\frac{(D-2)!}{(D-2m)!} \sqrt{-\gamma}\,r^{D-2m-2}
\left[1-(\nabla r)^2\right]^{m-2} \nonumber \\ &&\times \bigg\{
2m(m-1)r^2\left[(\nabla_a\nabla_br)^2-(\nabla^2r)^2\right] \nonumber\\
&&\quad+2m(D-2m)r\nabla^2r\left[1-(\nabla r)^2\right]
-m\mathcal{R}r^2\left[1-(\nabla r)^2\right] \nonumber \\
&&\quad\left. -(D-2m)(D-2m-1)\left[1-(\nabla r)^2\right]^2\right\}\;.
\end{eqnarray}
where $\mathcal{R}$ is a two-dimensional Ricci scalar calculated 
from $\gamma_{ab}$. Specializing further to $AdS_2\times S^{D-2}$ metric
(\ref{horfie}) all terms having covariant derivatives vanish on the
horizon and using this and (\ref{horfie}) one obtains the following 
expression for the function $f$
\begin{eqnarray}\label{fgb}
f &=& \frac{\Omega_{D-2}}{16\pi G_N}\,u_S\,v_1\,v_2^{(D-2)/2} \left\{ 
\frac{2\,u_T^2\,e_1^2}{v_1^2} + \frac{2\,e_2^2}{u_T^2\,v_1^2} \right. 
\\ \nonumber && + \sum_{m=1}^{[D/2]} 
\alpha'^{m-1} \lambda_m \frac{(D-2)!}{(D-2m)!} \left.
v_2^{-m} \left[(D-2m)(D-2m-1) - 2m\frac{v_2}{v_1} \right] \right\}
\end{eqnarray}
where $\lambda_1=1$.

Now we can use (\ref{eom}--\ref{ent}) to calculate entropy. For
better understanding we specialize first to $D\le7$ and then take
the general case.

\subsection{$D=4,5$}

In this case we have only $m=1,2$ terms in (\ref{fgb}). Although the 
analysis was already done in \cite{sen0505}, for completeness we 
shall repeat it here. From (\ref{fgb}) we get 
\begin{equation}\label{fgb45}
f = \frac{\Omega_{D-2}}{16\pi G_N}\,u_S\,v_1\,v_2^{(D-2)/2} \left[
\frac{2\,u_T^2\,e_1^2}{v_1^2} + \frac{2\,e_2^2}{u_T^2\,v_1^2} 
- \frac{2}{v_1} + \frac{(D-2)(D-3)}{v_2} \left( 1 -
\frac{4\,\alpha' \lambda_2}{v_1} \right) \right]
\end{equation}
Now we impose EOM's (\ref{eom}), and use (\ref{charge},\ref{q-nw}) to
express results in terms of $n$ and $w$. One obtains a unique solution 
\begin{eqnarray}
v_1 &=& 4\,\alpha' \lambda_2 \label{v145} \\
v_2 &=& 2(D-2)(D-3) \alpha' \lambda_2 \\
u_T &=& \sqrt{\frac{n}{w}}  \label{uT45} \\
u_S &=& \frac{4\pi G_N}{\Omega_{D-2}} \frac{v_1}{v_2^{(D-2)/2}}
\frac{q_1}{e_2} = \frac{4\pi G_N}{\Omega_{D-2}} 
\frac{v_1}{v_2^{(D-2)/2}}
\frac{\sqrt{2nw}}{\alpha'\sqrt{\lambda_2}} \label{uS45} \\
e_1 &=& \sqrt{2\,\alpha' \lambda_2 \frac{w}{n}} \;,\qquad\qquad
e_2 = \sqrt{2\,\alpha' \lambda_2 \frac{n}{w}} \label{e1245}
\end{eqnarray}
Using (\ref{v145}-\ref{e1245}) and (\ref{q-nw}) in (\ref{ent}) we
obtain the entropy
\begin{equation}\label{ent45l}
S = 4\pi \sqrt{8\,\lambda_2} \sqrt{nw}
\end{equation}
We now see that to match the statistical entropy of string states
(\ref{entropy}) one has to take
\begin{equation}\label{lam2}
\lambda_2 = \frac{1}{8}
\end{equation}
As noticed in \cite{sen0505} this is exactly the value which appears
in front of the Gauss-Bonnet term in the low energy effective action 
of heterotic strings. Observe also that by fixing only one parameter
$\lambda_2$ one obtains (\ref{entropy}) for both $D=4$ and $D=5$.

Notice here some aspects of solution which we shall show to be common
for all $D$. First, dilaton field $u_S\propto\sqrt{nw}$, so for the
effective string coupling on the horizon  $g^2=1/u_S\propto
1/\sqrt{nw}\ll 1$ for $n,w\gg1$. So, tree level in string coupling is
a good approximation. Second, $v_i\propto\alpha'$, which means that
all terms in our effective action are of the same order in $\alpha'$.
All higher curvature terms ar a priori important.

\subsection{$D=6,7$}

When we go up to $D=6$ and $D=7$, we see from (\ref{fgb}) that the 
function $f$ receives additional contribution (comparing to
(\ref{fgb45})), given by
\begin{equation}\label{fgb67}
\Delta f_{6,7} = \frac{\Omega_{D-2}}{16\pi G_N}\,u_S\,v_1\,v_2^{(D-2)/2}
(D-2)(D-3)(D-4)(D-5) \frac{\alpha'}{v_2^2} \left( \lambda_2 -
\frac{6\,\alpha' \lambda_3}{v_1} \right)
\end{equation}
We saw in the previous subsection that $\lambda_2=1/8$.

Now we solve the EOM's. It is obvious that we again obtain
(\ref{uT45}) and the first equality in (\ref{uS45}). Solving EOM's for
$v_1$ and $v_2$ we obtain
\begin{equation}
t_1 = \frac{t_2^2 + a(t_2+48b\lambda_3)}{a (t_2-8b)}
\end{equation}
where $t_2$ is a solution of the cubic equation
\begin{equation}\label{cubic}
t_2^3 - (a-b) t_2^2 - 144ab\lambda_3 t_2 - 48ab^2\lambda_3 = 0
\end{equation}
In the above formulae we have used the notation
\begin{equation}
t_i \equiv \frac{4v_i}{\alpha'}\;,\qquad a \equiv (D-2)(D-3)\;,\qquad
b \equiv (D-4)(D-5)
\end{equation}
For any given $\lambda_3$ we have generally three solutions for 
$v_{1,2}$, but it can be shown that there is only one physicaly
interesting for which both $v_1,v_2$ are real and positive. Using this
solution one can proceed further and as in $D=4,5$ solve all EOM's and 
calculate the entropy. As the corresponding expressions are cumbersome
and noniluminating functions of $\lambda_3$, we shall not write them 
explicitely.

The entropy (\ref{ent}) has the form
\begin{equation}\label{entD}
S = \omega(\lambda_3,D) \sqrt{nw}
\end{equation}
where $\omega$ is some complicated function of $\lambda_3$ and $D$.
Now, we search for such $\lambda_3$ for which in $D=6$ and
$D=7$ we obtain (\ref{entropy}). One way to fix $\lambda_3$ is to
demand\footnote{Equivalently, we could ask that $\omega=4\pi$ for
$D=6$, and then check do we obtain the same result for $D=7$.} that 
entropy is the same in both dimensions
\begin{equation}
\omega(\lambda_3,D=6) = \omega(\lambda_3,D=7)
\end{equation}
It is easy to show that the only solution is
\begin{equation}\label{lam3}
\lambda_3 = \frac{1}{96}
\end{equation}
Now we use this value for $\lambda_3$ in (\ref{entD}) and obtain that
the entropy is given by
\begin{equation}
S = 4\pi\sqrt{nw}
\end{equation}
which is again exactly the string result (\ref{entropy}). For the
choice (\ref{lam3}) solution is given by
\begin{eqnarray}
v_1 &=& \frac{\alpha'}{2} \label{v167} \\
v_2 &=& \frac{\alpha'}{8} (D-2)(D-3) \left[ 1 + 
\sqrt{1+\frac{2(D-4)(D-5)}{(D-2)(D-3)}} \,\right] \label{v267} \\
u_T &=& \sqrt{\frac{n}{w}} \label{uT67} \\
u_S &=& \frac{4\pi G_N}{\Omega_{D-2}} \frac{v_1}{v_2^{(D-2)/2}}
\frac{q_1}{e_2} = \frac{8\pi G_N}{\Omega_{D-2}}
\frac{\sqrt{nw}}{v_2^{(D-2)/2}} \label{uS67} \\
e_1 &=& \sqrt{\frac{\alpha'}{4} \frac{w}{n}} \;,\qquad\qquad
e_2 = \sqrt{\frac{\alpha'}{4} \frac{n}{w}} \label{e1267}
\end{eqnarray}

\subsection{General dimensions}

We now pass to general number of dimensions $D$ recursively. From 
(\ref{fgb}) we see that passing from (odd) dimension $D=2m-1$ to 
$D=2m$ and $D=2m+1$ the function $f$ gets additional contribution
\begin{equation}
\Delta f = \frac{\Omega_{D-2}}{16\pi G_N}\,u_S\,v_1\,v_2^{(D-2)/2} 
\alpha'^{m-2} \frac{(D-2)!}{(D-2m)!} v_2^{-m+1} 
\left( \lambda_{m-1} - \frac{2m\alpha'}{v_1} \lambda_m \right)
\end{equation}
We assume that all $\lambda_{k}$, $k=1,\ldots,m-1$ are determined
from lower-dimensional analyses, so the only free parameter at the
moment is $\lambda_m$.

In principle we could apply the same analysis as in previos
subsections, i.e., solve the EOM's, calculate the entropy for
general $\lambda_m$ and then look is there a value of $\lambda_m$ for
which the entropy is equal to (\ref{entropy}). The problem is that 
for this one has to solve polinomial equation, like (\ref{cubic}), 
which is now of the order $(2m-3)$ and so for $m\ge4$ cannot be 
solved analyticaly for general $\lambda_m$.

However, closer inspection of the solution (\ref{v167}-\ref{e1267}) 
for $D\le7$ reveals the shortcut. We notice that only $v_2$ depends on
$D$, and that $v_1$, $u_T$, $e_i$ are depending just on $n$ and $w$. 
From (\ref{q-nw}) and (\ref{ent}) we see that to obtain for the
entropy string result (\ref{entropy}) it is necessary that $e_i$ are
given by (\ref{e1267}). One obvious way to have this is to fix
$m\lambda_m/\lambda_{m-1}$ to be the same for all $m$. Then 
\begin{equation}\label{v1gen}
v_1 = 2m\alpha'\frac{\lambda_m}{\lambda_{m-1}}
\end{equation}
is one solution of EOM. Then, to have (\ref{e1267}) we see that $v_1$
has to be given by (\ref{v167}), which combined with (\ref{v1gen})
gives the coupling constants
\begin{equation}\label{coup}
\lambda_m = \frac{\lambda_{m-1}}{4m} = \frac{4}{4^m m!}
\end{equation}
where we have used $\lambda_1=1$.

To summarise, for the choice of coupling constants given in
(\ref{coup}) there is a solution\footnote{We
have checked that for $D\le9$ this is a unique solution with both 
$v_1$, $v_2$ real and positive.} of EOM for any $D$ given by
(\ref{v167}), (\ref{uT67}-\ref{e1267}), and with $v_2=\alpha'y(D)$,
where $y(D)$ is some complicated function of $D$ (which is a real 
and positive root of $(m-1)$-th order polynomial), for which the 
Wald entropy formula gives
\begin{equation}
S=4\pi\sqrt{nw} \;.
\end{equation}
And this is exactly the statistical entropy of half-BPS states of
heterotic string given in (\ref{entropy}).

\section{Some remarks}

Before discussing our results, let us make two remarks. First, we
would like to note that the gravitational part of the Lovelock type 
action with coefficients given by (\ref{coup}) apparently can be 
written in the exponential form
\begin{equation} \label{llexp}
S_{grav} = \frac{1}{4\pi G_N \alpha'} \int d^Dx \sqrt{-g} \,S
\left[ \exp \left( \sum_{m=1} \frac{\alpha'^m}{4} \lambda_m 
\tilde{\mathcal{L}}^{GB}_m \right) - 1 \right]
\end{equation}
where $\tilde{\mathcal{L}}^{GB}_m$ are obtained from the extended 
Gauss-Bonnet densities $\mathcal{L}^{GB}_m$ given in (\ref{lgbm}) by 
throwing away all terms which are products of two or more scalars 
(like e.g., $R^2$, $R(R_{\mu\nu})^2$, etc.). We do not have a proof of
this, but we have checked it explicitely for terms up to $\alpha'^3$ 
order, and also confirmed that terms of the type $R^k X$ are in 
agreement with the known recursion relation
\begin{equation}
\frac{\partial \mathcal{L}^{GB}_m}{\partial R} = 
m \mathcal{L}^{GB}_{m-1} \;.
\end{equation} 
This makes us believe that (\ref{llexp}) is correct. As far as we
know, the Lovelock action with the particular choice of parameters
given in (\ref{coup}) was not mentioned in the literature before.

For a second remark, notice that from (\ref{fgb}) and (\ref{coup}) 
follows that $f$ function can be put in the form
\begin{equation} \label{actgen}
f = \frac{\Omega_{D-2}}{16\pi G_N}\,u_S\,v_1\,v_2^{(D-2)/2} \left[
\frac{2\,u_T^2\,e_1^2}{v_1^2} + \frac{2\,e_2^2}{u_T^2\,v_1^2}
- \frac{2}{v_1} 
- \left(\frac{1}{v_1}-\frac{2}{\alpha'}\right) A \right]
\end{equation}
where the function $A$ is given by
\begin{equation}
A = A(v_2) = \sum_{m=1}^{[D/2]} \alpha'^m \lambda_{m+1} 
\frac{2m(D-2)!}{(D-2m-2)!} \frac{1}{v_2^m}
\end{equation}
Equation for $v_2$ ($\partial f/\partial v_2=0$) gives directly a
solution $v_1=\alpha'/2$, which substituted back into $f$ leaves just 
the term
\begin{equation}
-\frac{2}{v_1} = R_{AdS_2}
\end{equation}
plus the terms with gauge fields. In equation for dilaton $u_S$ 
(equivalent to $f=0$) all dependence on $v_2$ vanishes and we obtain
\begin{equation}
e_1 e_2 = \frac{\alpha'}{4}
\end{equation}
from which, using (\ref{ent}), we obtain result (\ref{entropy}) for the 
entropy without ever needing to solve for $v_2$.

It is obvious that in the arguments above a precise form of the 
function $A$ was completely arbitrary, moreover it could depend also on
$v_1$ and $e_i$. One always gets 
(\ref{v167},\ref{uT67}-\ref{e1267},\ref{entropy}) where the exact form
of $A(v_1,v_2)$ only affects the solution for $v_2$ (which affects 
also dilaton $u_S$ through (\ref{uS67}.). As a consequence, any action
which for the $AdS_2\times S^{D-2}$ near horizon geometry has the form
(\ref{actgen}) will give the same result for the entropy of 2-charged
black holes, i.e., (\ref{entropy}).

The same conclusion does not hold for 4-charged and 8-charged black
holes in $D=4$. In these cases there is additional term inside the 
square brackets in (\ref{actgen}) proportional to $v_2^{-2}$
\cite{sen0508} and only for some special choices of the function $A$ 
one would get the entropy equal to statistical entropy of string 
states.

\section{Discussion}

We have analysed solutions with $AdS_2\times S^{D-2}$ geometry in the
theories with actions of the Lovelock type which contain all 
extended Gauss-Bonnet densities. We expect that these solutions 
describe $D$-dimensional asymptoticaly flat two-charge black holes 
near the horizon. The idea was to check could Sen's results for
$D=4,5$ \cite{sen0505} be generalized to all dimensions.

In the lowest order in $\alpha'$ and string these actions are equal 
to truncated tree level (in string coupling {\em and} tension 
$\alpha'$) low 
energy effective actions of the heterotic string compactified on 
$T^{9-D}\times S^1$, for which analytic black hole solutions having 
singular horizon with vanishing area, and thus also the entropy, were
found \cite{sen9411,peet95}. They are believed to correspond to 
perturbative half-BPS states of heterotic string, for which the 
statistical entropy (i.e., logarithm of the number of states) is 
asymptoticaly given by (\ref{entropy}) \cite{DabHar89,DaGiHaRR90}. 
A reason for the discrepancy in the results for the entropy is that
these black holes are small, in fact singular, with the curvature 
diverging on the horizon. This suggests that higher curvature terms
in the action are important. On the other hand, dilaton field near the
horizon is large, which means that string coupling is small. One
concludes that it is necessary to consider effective action which is 
tree level in string coupling, but {\em not} in $\alpha'$.

Now, the small black holes we have analysed in this paper are
obviously some deformations of these singular black hole
solutions, but of course the question is have they anything at all 
with the black holes of heterotic string. We have shown that
parameters which appear in the Lovelock type action can be uniquely 
chosen such that the black hole entropy matches statistical entropy
of heterotic string states for {\em all} $D$. Moreover, this choice 
is nontrivial, in the sence that there is ``one parameter for every 
couple of dimensions''. Certainly, this matching could be just a 
coincidence. But, recently it was shown \cite{sen0508} that the same 
type of the action applied to 4-charge and 8-charge black holes in 
$D=4$ produced the same results for the entropy, gauge field strengths
and the axion-dilaton field as in the analyses based on supersymmetric
action obtained by supersymmetrizing square of the Weyl tensor
\cite{CaWiMo9906,CaWiKaMo04}.
Unfortunately, as corresponding supersymmetric formulations in $D>4$
are unknown it is impossible to make simmilar comparison in our case.
In spite of this, these results are hinting that there could be some
connection between the Lovelock type action we used and the heterotic 
string on the tree level in the string coupling. If true, then our
analysis shows how increasing the dimension $D$ naturally introduces 
terms of higher and higher order in curvature ($[D/2]$-order in $D$
dimensions).

Obviously, the action we used differs from the low energy
effective action of heterotic string on $M_D\times T^{9-D}\times S^1$
background. Alhough we do not know the exact form of the latter, we do
know that it should be supersymmetric and to contain additional
higher curvature terms beside extended Gauss-Bonnet ones, and also
higher derivative terms including gauge fields. Moreover,
it is known that $\mathcal{L}^{GB}_3$ term is not present in the low
energy effective action, and that some of the terms on the $m=4$ level
are proportional to the transcedental number $\zeta(3)$. This is in
contrast to our results $\lambda_3=1/96$ and $\lambda_4=1/3\cdot 2^9$.
On the other hand, as noted in \cite{sen0505}, the result
$\lambda_2=1/8$ is exactly the value which appears in the
low energy effective action of heterotic string
\cite{MetTse87,GroSlo87}. Curiously, $\lambda_3=1/96$ is
exactly the value which appears in the case of the bosonic string.
Here the following observation is important. Any term which is 
obtained by multiplying and contracting $m$ Riemann and field strength 
tensors evaluated on $AdS_2\times S^{D-2}$ background (\ref{horfie}) 
gives just a linear combination of terms $v_1^{-k}v_2^{k-m}$, 
$k=1,\ldots,m$ with some coefficients generally depending on $D$. 
Now, there is an infinite set of actions which are equivalent to ours
when evaluated on this background, and even bigger one consisting of
actions which lead to the more general form (\ref{actgen}). It can be
explicitely shown that one can use above this freedom to avoid 
disagreement with cubic and quartic higher curvature terms mentioned 
above. The question can supersymmetry be accomodated is opened. We 
shall present details elsewhere (\cite{PP2}).

It is clear that the sole results from this paper and from 
\cite{sen0505,sen0508} are insufficient for making any
strong claims. One can construct other actions leading to same 
results. As an illustration, let us consider an action obtained by 
adding higher curvature correction
\begin{equation}
\mathcal{L}_2 = \frac{1}{8} 
\left[ (R_{\mu\nu\rho\sigma})^2 - (R_{\mu\nu})^2 \right]
\end{equation}
to the leading term given by (\ref{alpha0}). This action does not
belong to the type (\ref{actgen}). It can be shown that it gives the
same result for the entropy (\ref{entropy}) as Lovelock type action 
for 2-charged black holes in $D=4$ and $D=5$, and for 4-charge and 
8-charge black holes in $D=4$. In fact, we could with this action
repeat the analysis in \cite{sen0508} and obtain exactly the same
solutions, including the atractor equations (4.11). Adding apropriate
higher derivative terms (with coefficients not depending on $D$) it 
is posible to match the entropy of 2-charge black holes in any
dimension $D$.

To conclude, the results in this paper support and extend to all
dimensions Sen's suggestion of a possible role of Gauss-Bonnet
densities in description of black holes in heterotic string theory.
It would be interesting to relate our results to the anomaly 
cancelation arguments of \cite{KraLar0506,KraLar0508}, especially 
concidering the topological origin of the extended Gauss-Bonnet 
densities. In any case, further analyses, including more examples, 
could either clarify this role, or to show that obtained agreement is
accidental.

\acknowledgments

I would like to thank J.\ Kappeli, A.\ Kleinschmidt, K.\ Peeters and
S.\ Theisen for valuable discussions. This work was supported by
Alexander von Humbold Foundation and by Ministry of Science, Education
and Sport of Republic of Croatia (contract No.\ 0119261).

\end{document}